\documentclass[aps,prl,twocolumn,groupedaddress,showpacs]{revtex4}

\usepackage{graphicx}
\usepackage{latexsym}

\begin{document}


\title{Precision measurement of gravity with cold atoms in
an optical lattice\\ and comparison with a classical
gravimeter.}


\author{N. Poli}
\author{F.-Y. Wang}\thanks{Also: ICTP, Trieste, I}
\author{M. G. Tarallo}
\author{A. Alberti}\thanks{Present address: Institut f\"{u}r Angewandte Physik der
Universit\"{a}t Bonn, Wegelerstrasse 8, 53115 Bonn, D}
\author{M. Prevedelli}\thanks{Permanent address: Dipartimento di Fisica, Universit\`a di Bologna, Via Irnerio
46, 40126 Bologna, I}
\author{G. M. Tino}\email{Guglielmo.Tino@fi.infn.it}

\affiliation{Dipartimento di Fisica e Astronomia and LENS, Universit\`a di
Firenze\\ INFN Sezione di Firenze, Via Sansone 1, 50019 Sesto
Fiorentino, Italy}


\date{\today}

\begin{abstract}
  We report on a high precision measurement of  gravitational acceleration
  using ultracold strontium atoms trapped in a vertical optical
  lattice. Using amplitude modulation of the lattice intensity, an uncertainty  $\Delta g /g \approx 10^{-7}$   was reached by measuring at the 5$^{th}$ harmonic of the
  Bloch oscillation frequency. After a careful analysis of systematic effects, the  value obtained with this microscopic quantum system is  consistent with the one we measured with a classical absolute gravimeter at the same location.
  This result is  of relevance for the recent interpretation of related experiments as tests of gravitational redshift and opens the way to new tests of gravity at micrometer scale.
\end{abstract}



\maketitle

Atom interferometry, and in general methods based on quantum
interference of ultracold atoms, were largely used in recent years
for gravitational physics experiments and new exciting prospects
can be envisioned in the near future \cite{Cronin2009}. For
example, Raman interferometry was used for precise measurements of
Earth's gravitational acceleration $g$ \cite{Peters99} and its gradient
\cite{McGuirk02}, for determining the value of the gravitational
constant \cite{Fixler07,Lamporesi08}, for a possible redefinition
of the kg \cite{Merlet08}, and for geophysical applications
\cite{DeAngelis09}. Schemes based on Bloch oscillations of atoms
trapped in vertical optical lattices were also used to measure
gravity with the possibility of combining high
sensitivity and micrometric spatial resolution
\cite{Clade2005,Ferrari06,Sorrentino09}. The results of atom interferometry
experiments were interpreted as tests of the isotropy of
post-Newtonian gravity \cite{Mueller2008}, of quantum gravity
\cite{Camelia2009}, and of gravitational redshift
\cite{Mueller2010}. Prospects include high precision tests of the
weak equivalence principle \cite{Fray04,Dimopoulos08a}, the detection
of gravitational waves \cite{Tino07b,Dimopoulos08b}, and future experiments in space
\cite{Tino2007}.

So far, however, Bloch oscillation measurements had limited accuracy compared to
Raman atom interferometers. Here, we present a precision measurement of gravitational
acceleration $g$ with a new method based on ultracold $^{88}$Sr atoms
trapped in an amplitude-modulated vertical optical lattice
\cite{Alberti2010} and compare the result with the value obtained
with a classical absolute gravimeter based on a Michelson interferometer with one arm
including a freely-falling corner-cube. We also improved our
previous observation of long-lived Bloch oscillations \cite{Ferrari06} and discuss
the precision of the two methods for the determination of $g$.
%
%
%
In addition to demonstrating the sensitivity and accuracy of this new
method, our data can be interpreted as a measurement of the gravitational
redshift   to the Compton frequency of Sr matter waves, as
suggested by H. M\"uller et al. \cite{Mueller2010}. Our data surpasses previous Bloch
oscillation measurements by one order of magnitude, making it the most precise test
of the gravitational redshift based on Bloch oscillations at micrometric spatial scales.
The interpretation of atom interferometer redshift
tests is complicated by special relativistic time dilation since the atoms
are moving \cite{Wolf2010,Mueller2010reply}, but Bloch oscillations experiments with stationary
lattices provide a measurement of the purely gravitational effect.

\begin{figure}[b]\begin{center}
\includegraphics[width=8 cm]{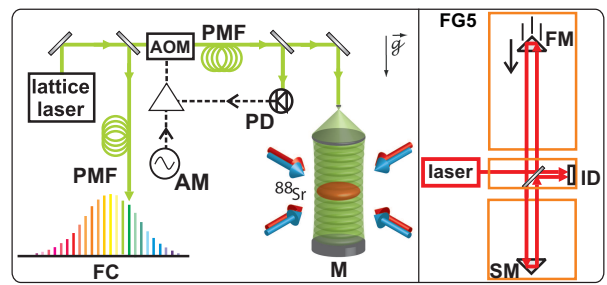}
\caption{Experimental setup for the measurement of gravity with $^{88}$Sr atoms
  trapped in a vertical optical lattice and the comparison with a classical absolute gravimeter (FG5). M: lattice mirror; FC: frequency comb; FM: freely-falling mirror; SM: stationary mirror; ID:
  interference detector; AOM: acousto-optical modulator;
  PMF: polarization maintaining fiber; AM: RF generator for amplitude modulation of the laser producing the lattice; PD:
  photodiode.\label{setup}}
\end{center}
\end{figure}

The experimental setup  is based on cooled
and trapped $^{88}$Sr atoms \cite{Ferrari06} (Fig.~\ref{setup}).  Atoms from a thermal beam are slowed in a
Zeeman slower and trapped in a ``blue'' Magneto Optical Trap (MOT)
operating on the $^1$S$_0$-$^1$P$_1$ resonance transition at
461~nm. The temperature is further reduced by a second cooling
stage in a ``red'' MOT operating on the $^1$S$_0$-$^3$P$_1$
intercombination transition at 689~nm. This produces about $10^6$
atoms at a temperature of 0.6~$\mu$K.
Since the force of gravity
is comparable to the force produced by the red MOT on the atoms,
the cloud of trapped atoms assumes a dish-like shape disk with a
vertical size of 27~$\mu$m and a radial size of 180~$\mu$m. The
atoms are adiabatically loaded in an optical lattice in
300~$\mu$s. The lattice potential is generated by a single--mode
frequency--doubled $\textrm{Nd:YVO}_4$ laser ($\lambda_L =
532$~nm) delivering up to 1~W on the atoms with a beam waist of
$557(7)~\mu$m. The beam is vertically aligned and retro-reflected
by a mirror.
The atomic sample in the lattice has a vertical RMS size of about
$14~\mu$m and a horizontal size of about
$100~\mu$m. The Bloch frequency is $\nu_B\simeq 574.3$~Hz
and the corresponding lattice photon recoil energy is $E_R \simeq
8\,\mathrm{kHz} \times h$. In typical conditions the lattice depth
ranges from 2.3 to 3~$E_R$, while the energy gap $E_G$ at the
recoil momentum $k_L$ is $E_G \simeq E_R$. The width of the first
energy band in the lattice potential is about $0.5 \times E_G$. Landau-Zener tunneling is  negligible in these conditions. The
lattice depth is stabilized by a servo loop acting on the
RF signal driving an acousto--optical modulator
(AOM). The same AOM is also used to add an amplitude modulation to
the lattice potential. The atomic cloud can be imaged either
\emph{in situ} or with usual time-of-flight technique using
resonant absorption imaging on a CCD camera with a spatial
resolution of 5~$\mu$m.
\begin{figure}[t]\begin{center}
\includegraphics[width=8 cm]{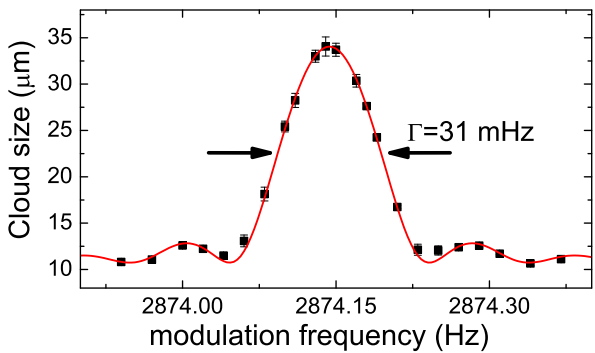}
\caption{
Spectrum recorded by  modulation of the lattice depth at the 5$^{th}$
harmonic of the Bloch frequency for 10.4 s
with a modulation depth of 7\%.  The red line is a fit of
      experimental data with a $\it{sinc}$ function.\label{Blochres}}
\end{center}
\end{figure}
The commercial
lattice laser (Coherent V--5) employed in the measurement is not
frequency stabilized and a precise calibration of its frequency is
then required.
To this purpose, part of the lattice laser light is
sent to a home--built self--referenced
Ti:Sa optical frequency comb.
Due to residual frequency instabilities on the
time scale of the Bloch frequency measurement, the
uncertainty in the laser frequency is $\simeq$ 100
MHz. 
The correction for the index of refraction of the Sr cloud
and the background gas in the vacuum chamber is negligible.
%
The vertical alignment of the lattice was checked with a precision of 0.5~mrad, corresponding to a relative uncertainty of 10 ppb on $g$, by overlapping the downward laser beam with the reflection from the surface of water
in a glass container inserted in the beam.
A tiltmeter with a
resolution of 1.7~$\mu$rad  attached to the optical table was employed to check the alignment stability during the measurements.

After  loading $^{88}$Sr atoms in the vertical lattice, the
trap depth is modulated sinusoidally. When the modulation frequency matches an integer harmonic of the Bloch frequency, atoms start to tunnel in neighbor
lattice sites giving rise to a net increase in the spatial
vertical atomic distribution which is observed \emph{in situ} by resonant
absorption imaging (see Fig.\ref{Blochres}).
The lifetime in the vertical lattice is about 20~s. This allows us
to apply amplitude modulation to the lattice for $\sim$10~s
resulting in an increased quality factor observed on the resonance
at $\nu_m=5\times\nu_B$. With a typical amplitude modulation depth
of the order of 7\% we estimate a tunneling rate
$\mathcal{J}_l/\hbar = 0.75$ \cite{Alberti2010}.
The recording time for a whole resonance spectrum is about 1~hour
and leads to a maximum resolution of 150~ppb in $\nu_B$. The value
of $g$ is given by  $g=2\,h\,\nu_B/(m_{\mathrm{Sr}}\lambda_L)$, where
$m_{\mathrm{Sr}}$ is the mass of $^{88}$Sr atoms and $h$ the
Planck constant which are both known with a relative
uncertainty of $\sim 5\times10^{-8}$.

\begin{table}[b] 
   \centering
   \begin{tabular}{@{} lcc @{}}
      \toprule
      Effect   & Correction  & Uncertainty \\
      \toprule
      Lattice wavelength     & 0 & 2 \\
      Lattice beam vertical align.& 0 &  0.1 \\
      Stark shift (beam geometry)& 14.3 $\div$ 17.3 & 0.4 \\
      Experiment timing  & 0 & 0.2\\
      Tides & -1.4 $\div$ 0.9 & $<$0.1\\
      Height difference & 4.3  &  0.2\\
      Refraction index & 0  &  $<$0.01 \\
      Fundamental constants   & 0  &  0.7 \\
      \hline
      Systematics total & 17.2 $\div$ 22.5  &  2.2\\
      \toprule
    \end{tabular}
    \caption{Systematic corrections and their associated uncertainties ($\times$ 10$^{-7}$)
for the gravity measurement with $^{88}$Sr atoms in the amplitude modulated optical lattice.}
    \label{syst}
  \end{table}
An important contribution to systematic shifts in  gravity
measurements with trapped neutral atoms is due to the lattice light
itself. Both the intensity and the wavevector of the lattice
beam, which results from the interference of two
counter-propagating Gaussian beams, yield space-dependent terms
to the potential  $U_{tot}(z) =
U_{s}(z)+U_{l}(z)\cos(2k(z)z)-mgz$, where $U_s(z)$ and $U_t(z)$
depend on the squared sum and on the product of the
two beam field amplitudes, respectively. This additional dependence of the
potential along the vertical direction gives rise to two correction
terms in the typical Bloch formula
$g=2\,h\,\nu_B/(m_{\mathrm{Sr}}\lambda_L)+\Delta g_U+\Delta g_k$
 given by the spatial derivative of the potential and
 the spatial derivative of the difference between the Gouy phase \cite{Clade'09} for the two beams
 at the position of the atomic cloud
$z_{at}$. The shift introduced by these two extra terms is
estimated by a precise determination of  the geometry of the
incoming and the reflected trapping beams and the position of the
cloud with respect to the beam waist, with a relative uncertainty
of 1\%. Moreover, an independent determination of the transverse
beam size at $z_{at}$ has been done by measuring the axial and
radial atomic trap oscillation frequencies through parametric
heating technique \cite{Savard97,Jauregui01}. For typical
experimental parameters, the two terms are $\Delta g_U $ =
1.53(3)x10$^{-5}$m/s$^2$ and $\Delta g_k$ =
1.0(2)x10$^{-8}$m/s$^2$.

Tidal effects were evaluated and removed from the raw
data using the same algorithm and potential model used for the
absolute gravimeter data processing. The peak--to--peak effect of tides
at our site is of the order of $2 \times 10^{-6} \mathrm{m/s^2}$. Since each measurement lasts
about 1 hour the variation of $g$ during a single measurement due
to tides is below $10^{-7} \mathrm{m/s^2}$ (i.e. below 10 ppb).
%
%
\begin{figure}[tt]
  \begin{center}
    \includegraphics[width=9 cm]{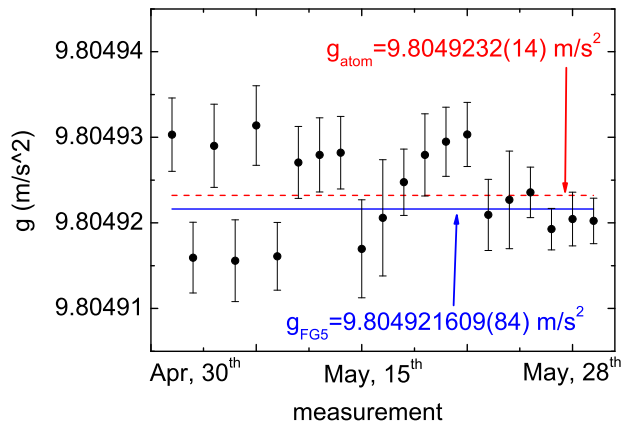}
    \caption{(color online).
      Measurements of $g$ using the amplitude modulation technique. Each
      experimental point is corrected for the systematic effects
      presented in Tab.~\ref{syst}. The red dashed line represents
      the weighted mean of the 21 measurements. The blue solid line is the value obtained with the classical absolute
      FG5 gravimeter.
      \label{meas}}
  \end{center}
\end{figure}
%

We checked also for a possible dependence from magnetic field
gradients by performing a set of measurements applying a
quadrupole magnetic field (from the MOT coils) up to
$B=$40~gauss/cm. The effect on $\nu_B$ is smaller than the statistical
error. All the other sources of systematic shift in the
measurement we evaluated (spurious higher harmonics of amplitude
modulation, Bloch-Siegert shift \cite{Bloch40}) are far below the
current accuracy level.
%
%
Table \ref{syst} summarizes the systematic shifts for the
gravity measurements with the
amplitude modulation technique. The values of individual shifts
depend on the experimental conditions; the quoted
uncertainties are typical values.

The reference value for local gravitational acceleration $g$ was provided by
an absolute gravimeter based on a Michelson interferometer with one arm
including a freely-falling corner-cube
 (FG5, Micro-g LaCoste). The measurement was performed in the same
laboratory at a distance of 1.15 m from the atomic probe position.
The difference in height of 14(5)~cm together with the estimated
vertical gravity gradient value $g_{zz}=-3.09 \times
10^{-6}\mathrm{s}^{-2}$ at the laboratory site was taken into
account in the data analysis. The result is
$g_{FG5}=9.804921609(84)$~m/s$^2$  \cite{DeAngelis10}.


Fig.~\ref{meas} presents a set of 21 determinations of $g$ with  $^{88}$Sr atoms.  The error bars are given by the quadrature
      sum of the statistical errors coming from the fit of the amplitude modulation resonance and
      the uncertainty on systematic corrections.  The standard deviation is $\sigma=110$~ppb with a
$\chi^2$=30. The resulting statistical
uncertainty  is $\sigma\times\sqrt{\chi^2/(n-1)}$=140~ppb.
The weighted mean of our data is $g_{Sr}=9.8049232(14)$~m/s$^2$,
in good agreement with the value obtained using the FG5 gravimeter.
%

\begin{figure}[t]
  \begin{center}
  \includegraphics[width=8 cm]{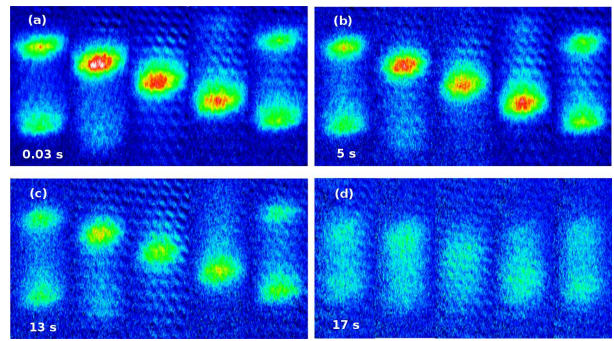}
    \caption{(color online). Long-lived Bloch oscillations for Sr atoms in the vertical lattice
    under the influence of gravity.
    Each picture shows one Bloch cycle in successive
time-of-flight absorption images giving the
momentum distribution at the time of release from the lattice.
Displayed are the first (a), the 2900th (b), the 7500th (c), and
the 9800th (d) Bloch cycle.
      \label{bloch}}
  \end{center}
\end{figure}

With minor modifications of the experimental procedure,  in this work we also determined $g$
by measuring the frequency of the Bloch oscillations of the atoms in the vertical optical lattice. Thanks to a better vacuum and taking advantage of the lattice modulation method to reduce the initial momentum distribution of the atoms in the lattice, we considerably improved the visibility and duration of the oscillations and, as a consequence, the frequency resolution compared  with previous experiments \cite{Ferrari06}.
 After the transfer of ultracold atoms in the
vertical optical lattice, an amplitude modulation burst with
typical duration of 120 cycles at $\nu_m\simeq\nu_B$ is
applied. The  quantum phase of the
atomic wavefunction induced by the  amplitude modulation gives rise
to an interference effect in time of flight image of the atomic
cloud which results in an enhanced visibility of the Bloch oscillations peaks~\cite{Alberti09}. After the modulation has turned off, we
let the atomic cloud evolve for a time T. Finally, we switch off
the confinement within 5~$\mu$s to measure the momentum
distribution taking an absorption picture on the CCD camera of the
atoms in ballistic expansion. In order to optimize the visibility
through this quantum interference effect we set the time-of-flight
to 14~ms.
As shown in Fig.~\ref{bloch}, we  observe  Bloch oscillations lasting up to 17 s. From the the  fit of the mean
atomic momentum we can estimate the Bloch frequency with 170~ppb
statistical uncertainty.
%
In comparison with the determination of Bloch
frequency obtained with the resonant amplitude modulation technique, however, we
observed a considerably larger scattering in repeated measurements, mainly
due to initial position instability of the atomic trap and also
higher dependence on the timing of the experiment.
The value for $g$  obtained with the
Bloch oscillation technique is indeed $g_{Bloch}=9.80488(6)$~m/s$^2$, which is
consistent with the measurement discussed above but is affected by a larger
uncertainty of 6~ppm.
%

It is important to notice that the amplitude modulation technique
for gravity measurement allows further improvements of both
accuracy and sensitivity. In fact our result is mainly limited
by  the lattice wavelength stability and by the Stark
shift. The first effect could be lowered by using a tunable laser and
locking it to an atomic line. For instance, if the wavelength is
stabilized within 1~MHz, the uncertainty of this effect might be
reduced by two orders of magnitude. The second main systematic
effect could be reduced either by using a blue-detuned trapping
laser~\cite{Clade'09} or by increasing the lattice beam waist.
Also, the sensitivity could be increased by using higher
harmonic amplitude modulation frequencies or by modulating for
a longer time.

In conclusion, we performed an accurate measurement of
 gravitational acceleration using  ultracold atoms trapped in a vertical optical lattice. The result is in  agreement at 140 ppb level with an independent determination obtained with a
classical FG5 gravimeter. This result represents
an improvement by an order of magnitude over previous results
\cite{Ferrari06,Ivanov08} and is of interest as a test
of the gravitational redshift~\cite{Mueller2010}. Moreover we
demonstrated the validity of the amplitude modulation
technique~\cite{Alberti2010} for the measurement of forces
with high spatial resolution~\cite{Sorrentino09}. We also
observed persistent Bloch oscillation up to 17 s which represents  the longest coherence
time observed to date. This result might also have important
applications in precision measurements in conjunction
with nondestructive cavity QED techniques to probe  atomic
momentum oscillations~\cite{Peden09}.

\begin{acknowledgments}
We thank ENI and INGV for the measurement with the FG5 gravimeter.
We also thank M. Schioppo for his contribution in the early stage of the experiment,  D. Sutyrin for help with absolute
frequency measurements and R. Ballerini, M. De Pas, M. Giuntini, A. Hajeb, A. Montori for technical assistance. This work is supported by INFN and LENS (under contract RII3 CT 2003 506350).


\end{acknowledgments}


%

\end{document}